\documentstyle[graphicx]{mn}

\newif\ifAMStwofonts
\AMStwofontstrue

\def\pg{{PG1211+143}}
\def\ngc{{NGC4051}}

\def\me{{\dot M_{\rm Edd}}}

\def\xmm{{\it XMM-Newton}}
\def\chandra{{\it Chandra}}
\def\suzaku{{\it Suzaku}}

\def\astroh{{\it Astro-H}}
\def\et{{et al.\ }}

\def\swift{{\it Swift}}

\newcommand{\ls}{\mathrel{\hbox{\rlap{\hbox{\lower4pt\hbox{$\sim$}}}\hbox{$<$}}}}
\newcommand{\gs}{\mathrel{\hbox{\rlap{\hbox{\lower4pt\hbox{$\sim$}}}\hbox{$>$}}}}

\def\Msun{\hbox{$\rm ~M_{\odot}$}}

\def\rchi{{$\chi^{2}_{\nu}$}}

\def\H0{{\rm ~km~s^{-1}~Mpc^{-1}}}

\def\et{{et al.}}

\def\deg{^\circ}

\title[Detection of a second high velocity]
        {Detection of a second high velocity component in the highly ionized wind from PG 1211+143}
\author[Ken Pounds \et]
        {Ken Pounds $^{1}$, Andrew Lobban $^{1}$, James Reeves $^{2}$ and Simon Vaughan $^{1}$ \\
$^{1}$ Department of Physics and Astronomy, University of Leicester, Leicester, LE1 7RH, UK \\
$^{2}$ School of Physical Sciences, Keele University, Keele, ST5 5BG, UK \\}

\date{Accepted ; Submitted }
\pagerange{\pageref{firstpage}--\pageref{lastpage}}
\pubyear{2015}
\begin{document}
\maketitle
\label{firstpage}

\begin{abstract}
An extended \xmm\ observation of the luminous narrow line Seyfert galaxy \pg\ in 2014 has revealed a more complex highly ionized, high velocity outflow. The detection of previously unresolved spectral
structure in Fe K absorption finds a second outflow velocity component of the highly ionized wind, with an outflow velocity of v$\sim$0.066$\pm$0.003c, in addition to a still higher velocity outflow of
v$\sim$0.129$\pm$0.002c consistent with that first seen in 2001. We note that chaotic accretion, consisting of many prograde and retrograde events, offers an intriguing explanation of the dual
velocity wind. In that context the persisting outflow velocities could relate to physically distinct orientations of the inner accretion flow, with prograde accretion yielding a higher launch
velocity than retrograde accretion in a ratio close to that observed.
\end{abstract}


\section{Introduction}

X-ray spectra from an \xmm\ observation of the luminous Seyfert galaxy \pg\ in 2001 provided the first detection in a non-BAL AGN of strongly blue-shifted absorption lines of highly ionized gas, 
corresponding to a sub-relativistic outflow velocity of $\sim$0.09c (Pounds \et\ 2003). That velocity was based primarily on the identification of a strong absorption line at $\sim$7 keV with
the resonance Lyman-$\alpha$ transition in Fe XXVI. The subsequent inclusion of additional absorption lines, of Ne, Mg, Si and S, led to the re-identification of the $\sim$7 keV absorption line with the resonance 1s-2p
transition in Fe XXV, and a revised outflow velocity of 0.14$\pm$0.01c (Pounds and Page 2006). Further observations of \pg\ over several years  with \xmm, \chandra\ and \suzaku\ found the high velocity
outflow to be persistent but of variable strength (Reeves \et\ 2008). Evidence that the mean outflow in \pg\ was both massive and energetic - with potential importance for galaxy feedback - was obtained from the
detection of P Cygni and other broad emission features by combining the 2001, 2004 and 2007 \xmm\ EPIC spectra (Pounds and Reeves 2009). 

Examination of archival data from \xmm\ and \suzaku\ subsequently showed ultra-fast, highly-ionized outflows (UFOs) to be relatively common in nearby, luminous AGN (Tombesi 2010, 2011; Gofford 2013). The
frequency of these detections appeared to confirm a substantial covering factor, with a persistent wind having sufficient mechanical energy to  disrupt the bulge gas in the host galaxy 
(Pounds 2014a). An indication 
how much of that
energy could be lost before reaching a still-active star forming region came with evidence of a UFO shocking against the ISM or slower-moving ejecta in the low mass Seyfert
galaxy \ngc\ (Pounds and Vaughan 2011,  Pounds and King 2013).  

Determination of the correct wind velocity, v, is critically important in estimating the mass and energy rates of AGN winds, with the latter being dependent on v$^{3}$ in a  radial outflow. However, significant
uncertainty remains when the deduced velocity depends on the identification of a single blue-shifted absorption line, typically of Fe K, as in many early UFO detections (eg Tombesi \et\ 2010). 
Spectral modelling over a wider energy band reduces ambiguities in line identification, as first demonstrated in Pounds and Page (2006), while also providing additional outflow parameters, including
the mean ionization parameter and column density. Such broad-band spectral modelling was used in the second paper on the \xmm\ archival data (Tombesi \et\ (2011) and by Gofford (2013) in their similar analysis of \suzaku\ data.

In the present paper we make a further addition, in simultaneously modelling the outflow of \pg\ with both 
photoionized absorption 
and  emission grids.    To quantify absorption and emission 
structure we employ photoionized grids of pre-computed spectra based on the xstar code (Kallman \et\ 1996), with the publicly available
grid 25 (turbulence velocity 200 km s$^{-1}$) being found to provide a satisfactory match to the highly-ionized absorption, and velocity-broadened emission adequately modelled by grid 22
(turbulence velocity 3000  km s$^{-1}$). 

Initially employing that procedure in re-modelling the 2001 outflow we re-affirm a column density N$_{H}$ of $\sim$$10^{23}$ cm$^{-2}$, ionization parameter log$\xi$$\sim$2.9 erg cm
s$^{-1}$ and outflow velocity v$\sim$0.14$\pm$0.01c. By comparison with the typical properties of UFOs (Tombesi \et\ 2011) and with the new observations reported here, it appears the unusually low outflow ionization
- and correspondingly high opacity - in the 2001 observation was a primary reason for it being amongst the most highly visible UFOs yet seen.

\section{A new \xmm\ observation of \pg\ in 2014}

In order to further explore the ionized wind in \pg\ an extended \xmm\ observation was carried out during 7 spacecraft orbits  over the period 2014 June 2 to 2014 July 9.  The present paper describes the
analysis of high energy spectral data, with subsequent papers reporting the detection of a low ionization outflow in the soft x-ray spectrum, and evidence for short-term variability. We use data
from the European Photon Imaging Cameras (EPIC) pn CCD (Strueder \et\ 2001), operated in large window mode.  

On-target exposures for individual orbits were typically $\sim$100 ks, apart from the fifth observation  (rev2664) of  $\sim$55 ks, giving a total on-target exposure of $\sim$630 ks. Full details of the \xmm\
observing log and data extraction procedures, and of accompanying \swift\ observations, are given in Lobban \et (2015), reporting the results of a detailed timing analysis. 

Raw data were processed using version 14.0 of the \xmm\ Scientific Analysis Software \footnote{http://xmm.esac.esa.int/sas/} package, following standard procedures.  Source events  were extracted from
20\,arcsec circular regions centred on the AGN, while background events were taken from much larger regions away from the target source and  from other nearby background sources.   For our EPIC-pn
analysis, we utilized single and double good pixel events ({\tt PATTERN} $\leq$ 4).   The exclusion of short intervals of
background flaring resulted in a final $\sim$570\,ks of high quality pn data.  The total residual background count rate was $<$1 per cent of  the source rate for all three EPIC cameras in
each observation (Figure 2).

\begin{figure}                                                          
\centering                                                              
\includegraphics[width=6.4cm, angle=270]{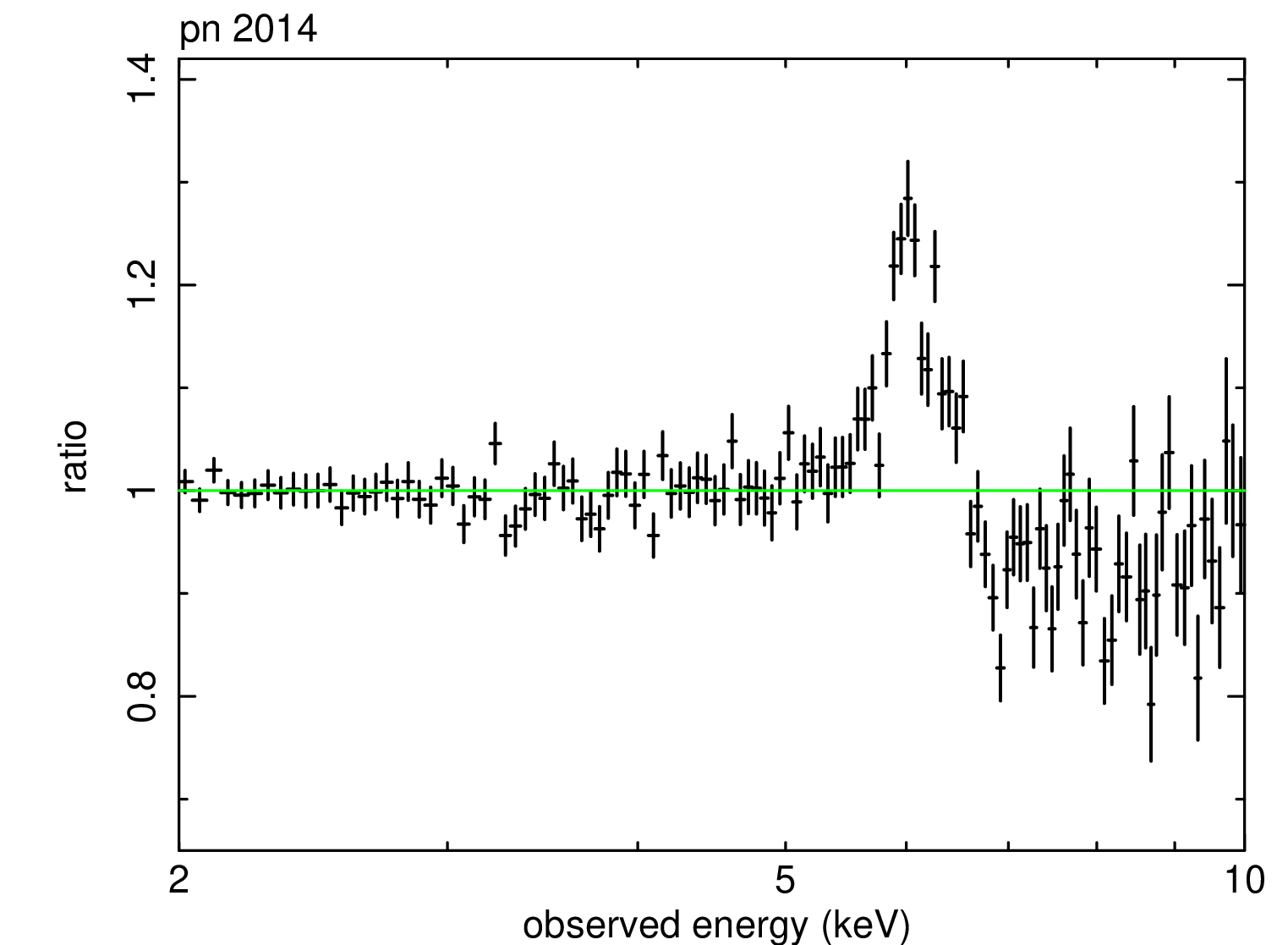}                                                 
\centering                                                              
\includegraphics[width=6.4cm, angle=270]{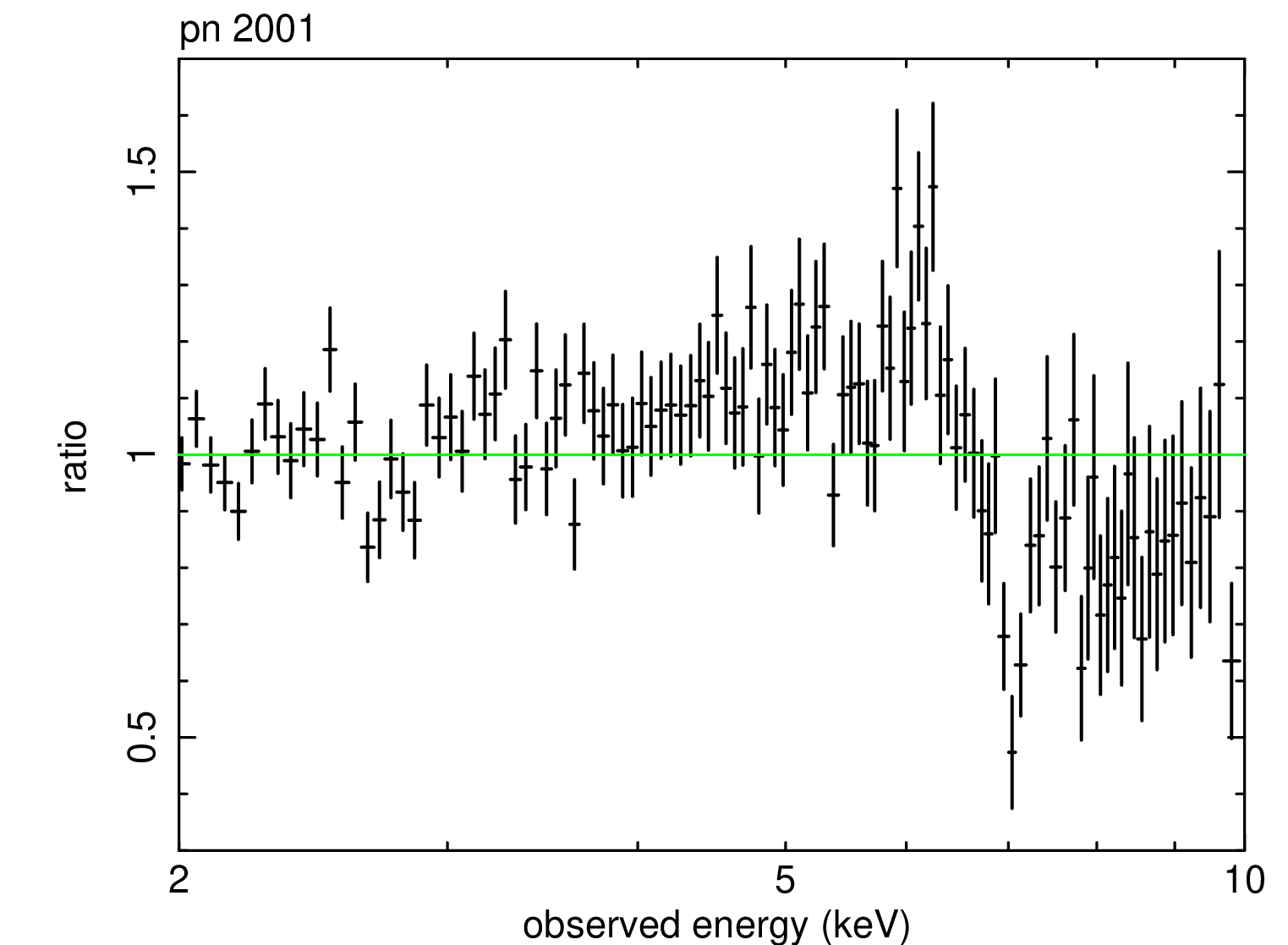}                                                 
\caption{Comparison of the stacked pn data from the 2014 observation of \pg\ with that from the 2001 observation, both plotted as a ratio to a double power-law continuum. While the most prominent absorption
line near 7 keV is weaker than in 2001, the much longer 2014 exposure allows the overall spectrum to be better defined. In particular, the Fe K emission profile is resolved into high and low ionization
components  and additional absorption structure is indicated at higher energies}  
\end{figure}

We assume an AGN redshift of $z=0.0809$ (Marziani \et\ 1996). Spectral modelling is based on the {\tt XSPEC} package (Arnaud 1996) and includes absorption  due to the line-of-sight Galactic column of
$N_{H}=2.85\times10^{20}\rm{cm}^{-2}$  (Murphy \et\ 1996). 90 \% confidence intervals on model parameters are based on $\Delta\chi^{2}=2.706$. Published estimates for the black hole mass in \pg\ range from
$3\times 10^{7}$\Msun\ (Kaspi \et\ 2000) to $1.5\times 10^{8}$\Msun\ (Bentz \et\ 2009), with the lower value making the mean luminosity close to the Eddington limit. All derived velocities are
corrected for the relativistic Doppler effect.

\section{Confirmation of a high velocity wind in \pg}

While examination of pn spectra from the 7 individual orbits in 2014 showed strong variability in the soft x-ray band, the harder continuum changes relatively little above $\sim$2 keV. To obtain  a baseline
hard x-ray spectrum of \pg\ with maximum statistical quality we therefore summed low background pn data from all 7 orbits for an initial  spectral analysis. The resulting low
background source exposure is a factor $\sim$10 greater than for any of the previous \xmm\ observations in 2001, 2004 and 2007.

Figure 1 compares the stacked pn spectrum from 2014 with that from 2001, both data sets being plotted as a ratio to a double power-law continuum of photon index $\Gamma$$\sim$1.6 and  
$\Gamma$$\sim$3 to match a variable `soft excess' at the lowest energies. Visual comparison of the two plots confirms that emission and
absorption  features are  significantly better defined  in the longer 2014 exposure. The emission  near 6 keV is now resolved into low and high ionization components, and the prominent absorption line at $\sim$7
keV is again seen in the 2014 data, albeit being less deep than in 2001. Other differences in the 2014 spectrum include weaker absorption in the lower mass ions, but a clear indication of additional absorption structure in
the Fe K band between 6-10 keV. Those differences suggest the fast wind in 2014 was more highly ionized than in 2001, a suggestion we now confirm by modelling the broad band spectrum from 2014 with the same
photoionized grids used to parameterise the 2001 outflow.

Figure 3 (top panel) shows a section of the stacked pn spectrum covering the Fe K region. The data have been grouped with a minimum of 25  counts (for $\chi^{2}$ compatibility) and with a maximum of 3
data points per pn camera resolution (FWHM) to optimise the visibility of spectral structure.  The emission  near 6 keV  is resolved into components  with rest energies close to the neutral Fe-K fluorescent
emission line and the 1s-2p resonance emission lines of He-like and H-like Fe, respectively.  The absorption line at $\sim$7 keV is a factor $\sim$3 weaker than the corresponding line in 2001, with an
equivalent width of 37$\pm$5 eV. In contrast, the deep 2014 exposure reveals additional structure at $\sim$6--10 keV, with apparent absorption lines at $\sim$6.9, $\sim$7.3, $\sim$7.5, $\sim$7.8, $\sim$8.2 and
$\sim$8.6 keV. \footnote {The absorption lines at $\sim$8.2 and $\sim$8.6 keV lie close to fluorescent x-ray emission lines of Cu and Zn arising from energetic particle impacts on the pn camera electronics
board (Strueder \et\ 2001). Fortunately, the low particle background throughout most of the 2014 \xmm\ observations (Figure 2) ensured such background  features have a negligible effect on the source
spectrum, an outcome confirmed by obtaining a very similar ratio plot to that in Figure 3 when no background data is subtracted.} 

Gaussian fitting of those spectral features is detailed in Section 4.1, and while the significance of individual lines is limited, we emphasise that finding a common velocity for a line series can be highly
significant, again underlining the importance of  taking proper account of multiple absorption lines in the study of x-ray outflows.   Analysis and interpretation of the Fe K absorption line structure, providing further detail on
the highly ionized wind in \pg, is the main purpose of this paper. 

\begin{figure}                                                          
\centering                                                              
\includegraphics[width=6.4cm, angle=270]{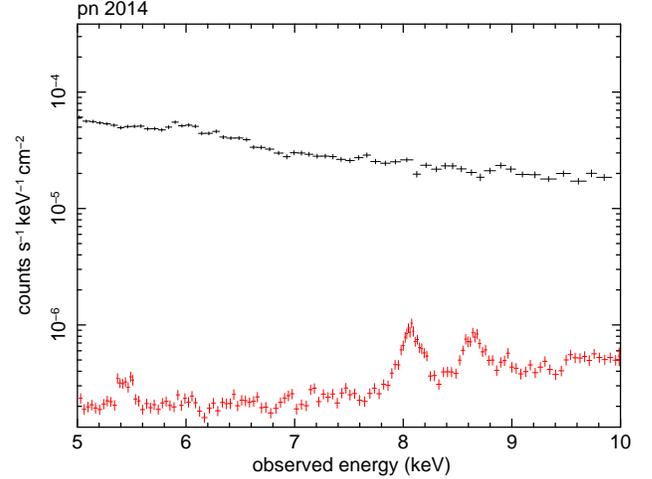}                                                                                                 
\caption{Comparison of the source (black) and background spectra (red) for the stacked 2014 pn data. The source spectrum is plotted without background subtraction and is near identical to that including background
subtraction, confirming the Fe K absorption features are 
not due to fluorescent Cu and Zn lines from the detector structure}  
\end{figure}

\begin{figure}                                                          
\centering                                                              
\includegraphics[width=6.4cm, angle=270]{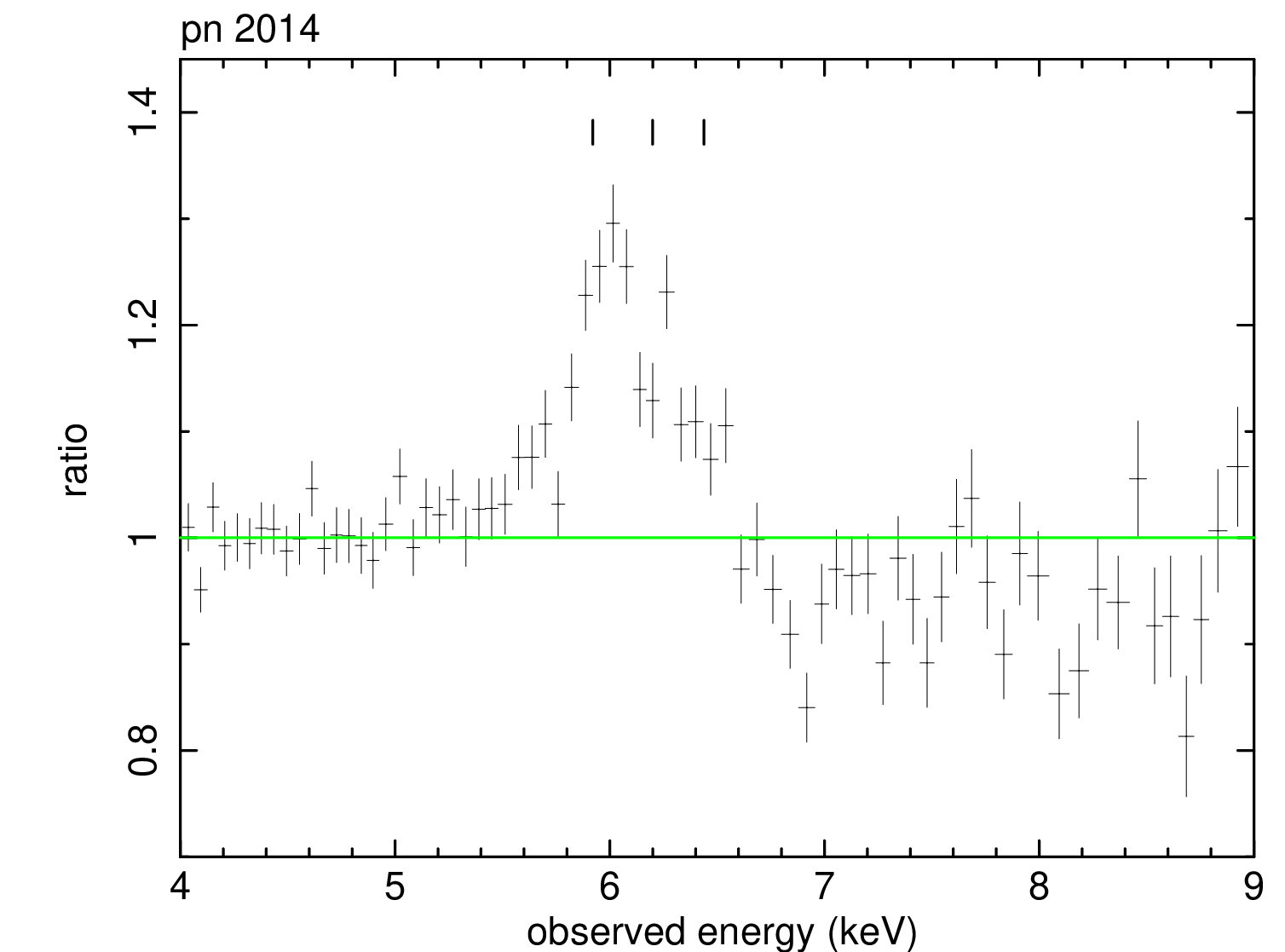}                                           
\centering                                                              
\includegraphics[width=6.0cm, angle=270]{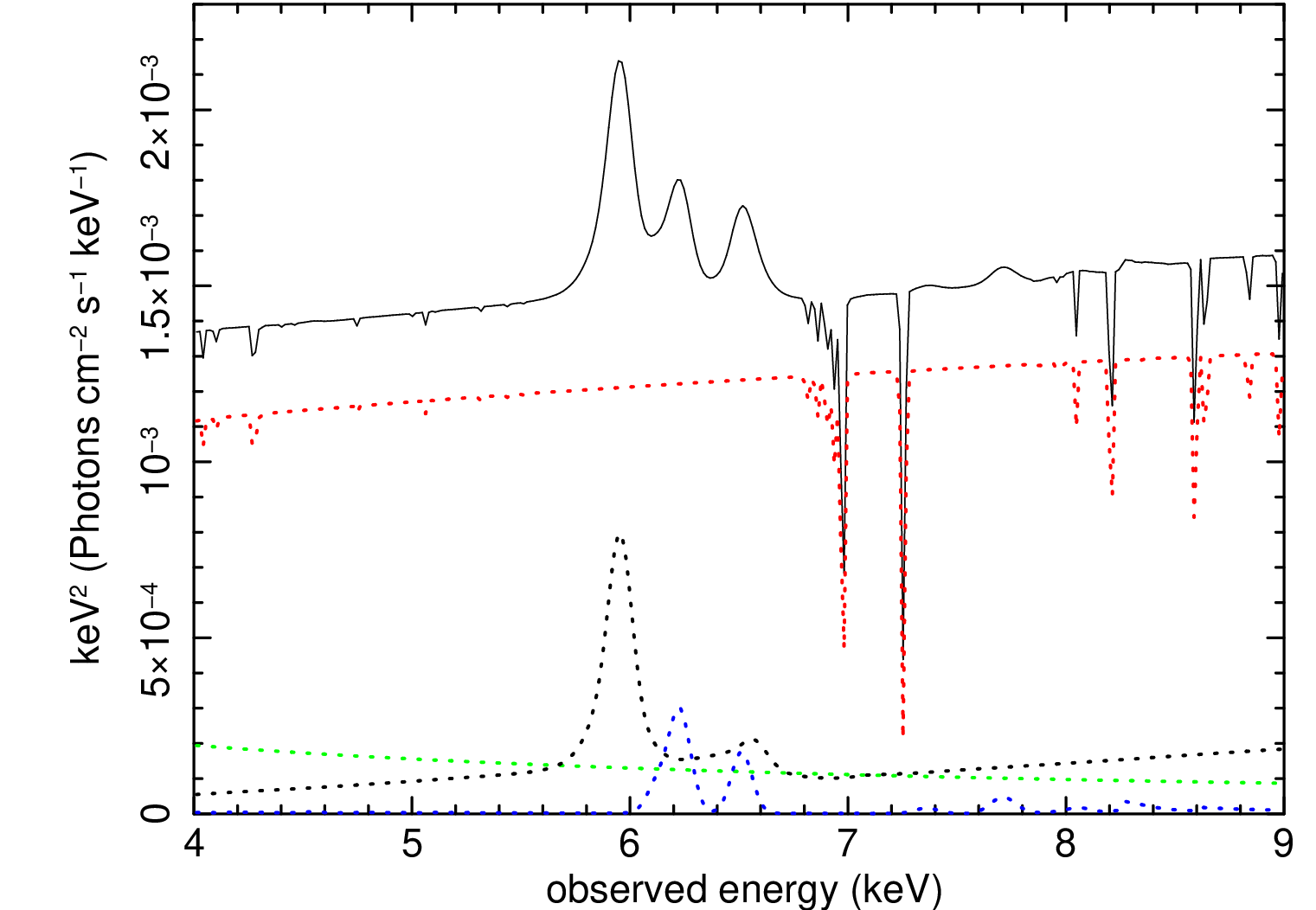}       
\caption{(top) Spectral structure in the stacked 2014 pn spectrum from Fig 1 resolves Fe K emission components corresponding to the Fe K fluorescence line and resonance emission from  
Fe XXV and Fe XXVI ions.  (lower) The addition of photoionized emission and absorption spectra match the resonance emission lines observed at $\sim$6.2 keV and  $\sim$6.45 keV, with photoionized absorption imprinted on the 
hard power law continuum (red).   A single outflow velocity identifies absorption lines at $\sim$7.0, $\sim$7.3, $\sim$8.2 and  $\sim$8.6 keV with the
$\alpha$ and $\beta$ lines, again of Fe XXV and XXVI. The model includes continuum reflection and linked Fe K fluorescence line emission (black) and is completed with a softer, unabsorbed power law (green) required by 
inter-orbit difference spectra (see text)}     
\end{figure}

Before testing for ionized absorption it is especially important to correctly model the underlying continuum for such high quality data. In particular, scattering from optically thick matter  (or so-called
'reflection' found to be common in AGN spectra (Nandra and Pounds 1994)) may impose a step change near the Fe K absorption edge  ($\sim$6.5 keV in observer space), and an indication of such an effect can be seen
in Figure 3 (top panel). To quantify that 'reflection' we added a {\sc Xillver} component (Garcia \et\ 2013) to the power law continuum model in XSPEC, simultaneously matching the fluorescent Fe K emission line and
continuum reflection from optically thick ionized matter. When previously modelled with a Gaussian, the Fe K line  energy of $\sim$5.96 keV ($\sim$6.44 keV at the AGN redshift), and EW of 73 eV yielded a 2--10 keV
spectral fit statistic of  (\rchi\ of 1513/1364). Re-fitting the Gaussian with {\sc Xillver} provided a corresponding adjustment to the continuum from reflection, with a further improvement in \rchi\ of 1495/1364.
The reflection component had an ionization parameter of log$\xi$$\sim$2.1, with inclination fixed at 45$\deg$ and a small outflow velocity  of 1500$\pm$1200 km s$^{-1}$  in the AGN rest-frame. Critically, although
reduced in depth, the absorption features at $\sim$7--9 keV in Fig.3 (top) remained clearly visible.  

Having modelled the underlying continuum, a photoionized absorber was then added to match that absorption structure, obtaining a significant improvement to the spectral fit (\rchi\ of 1470/1361), for a column density N$_{H}$ of 7.2$\pm$$5.0\times10^{22}$ cm$^{-2}$, ionization parameter
log$\xi$=3.3$\pm$0.1 erg cm s$^{-1}$ and outflow velocity of 0.119$\pm$0.003c. The addition of a photoionized emission spectrum, with tied ionization parameter, reproduced the ionized emission lines at $\sim$6.2 and
$\sim$6.45 keV, and further improved the fit (\rchi\ of 1445/1359), with the column density of the absorber decreasing to N$_{H}$ of 5.8$\pm$$3.2\times10^{22}$ cm$^{-2}$. The tied ionization parameter (log$\xi$=3.42$\pm$0.05 erg cm s$^{-1}$) and absorber velocity were unchanged in this second fit, while untying the ionization parameters of the emission and absorption spectra made little
difference to $\chi^{2}$. The ionized emission spectrum was found to have a blueshift (relative to the AGN), corresponding to an outflow velocity of 4200$\pm$1200 km s$^{-1}$, which we take to represent a mean
value averaged over an extended ionized outflow.  

Figure 3 (lower panel) illustrates the key 4-9 keV section of the single velocity outflow model, with absorption lines seen in the data at $\sim$7 and $\sim$7.3 keV identified with the 1s-2p resonance lines of FeXXV and XXVI, and absorption lines at
$\sim$8.2 and $\sim$8.6 keV with He-$\beta$ and a blend of Lyman-$\beta$ and He-$\gamma$ of the same ions. Finding a match to four absorption features with the same outflow velocity adds confidence in the robustness of the
spectral fit. 

The photoionized emission spectrum in Figure 3 correspondingly matches the two high energy components in the ratio plot, being identified with the He- and H-like resonance lines, in a ratio set by the  linked ionization
parameter. Encouragingly, the He-$\beta$ emission line can also be seen at $\sim$7.7 keV in both data and model, with a similar blueshift.

\subsection{A second velocity component in the highly ionized wind of \pg.}

Examination of Figure 3 shows that while the absorption lines seen at $\sim$7, $\sim$7.3, $\sim$8.2 and $\sim$8.6 keV are matched by resonance and higher order transitions in Fe XXV and Fe XXVI in the photoionized
outflow, the comparison of model and data is incomplete. To seek a further improvement in the fit a second absorber was therefore added, with ionization parameter, column density and velocity again free. Re-fitting the 2-10 keV
spectrum showed a further substantial improvement to the fit  (\rchi\ of 1415/1356), with the second absorption component having a significantly lower outflow velocity of
0.066$\pm$0.003c.

Figure 4 (top) illustrates the dual velocity absorber model, with the Lyman-$\alpha$ line of the lower velocity flow now blending with  He-$\alpha$ from the high velocity wind to provide a better match to  the broad absorption
feature observed at $\sim$6.9 keV. A consequence of that line blend is to increase the relative strength of the higher velocity Lyman-$\alpha$ line and hence - substantially - both the ionization parameter and column density of that
flow component. The higher outflow velocity is also increased slightly in the dual velocity fit, to 0.129$\pm$0.002c. 

An important consequence of simultaneously modelling the absorption and (re-)emission spectra is demonstrated by the He-$\alpha$ absorption line in the lower velocity outflow, previously partly hidden by the
emission line at  $\sim$6.5 keV. In turn, the partial absorption of the Fe Lyman-$\alpha$ emission line explains the relative weakness of that emission component in the data ratio plot.  
Comparing Figures 3 and 4, and the fit residuals, confirms that a significant contribution to the improved dual absorber fit lies in better matching the observed opacity at $\sim$7 keV, while  the lower
velocity  He-like absorption line models the sharp drop in the data at $\sim$6.6 keV as it cuts into the blue wing of the H-like emission line. 

\begin{figure}                                                                                                        
\centering                                                              
\includegraphics[width=6.4cm, angle=270]{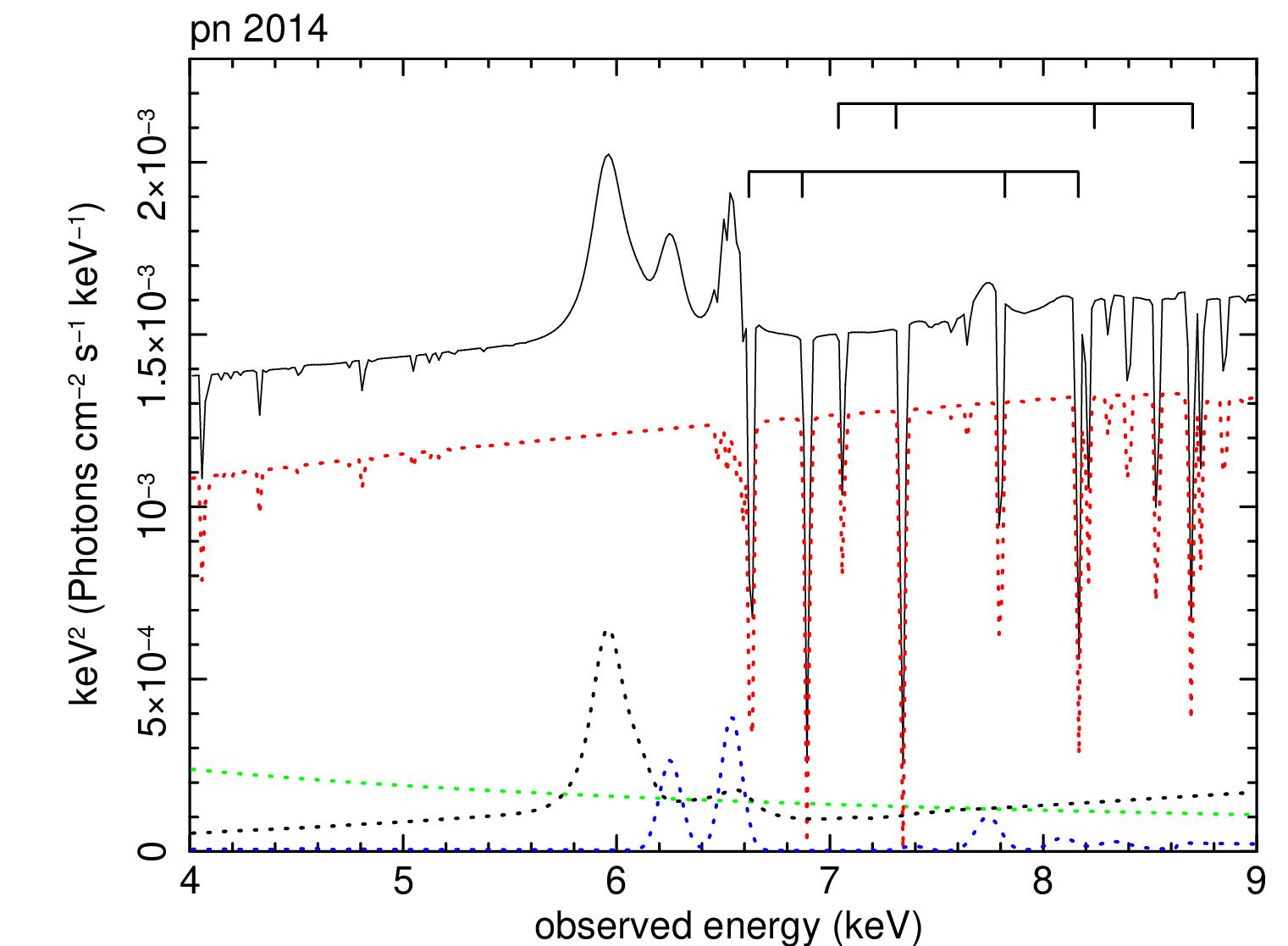}  
\centering                                                              
\includegraphics[width=6.0cm, angle=270]{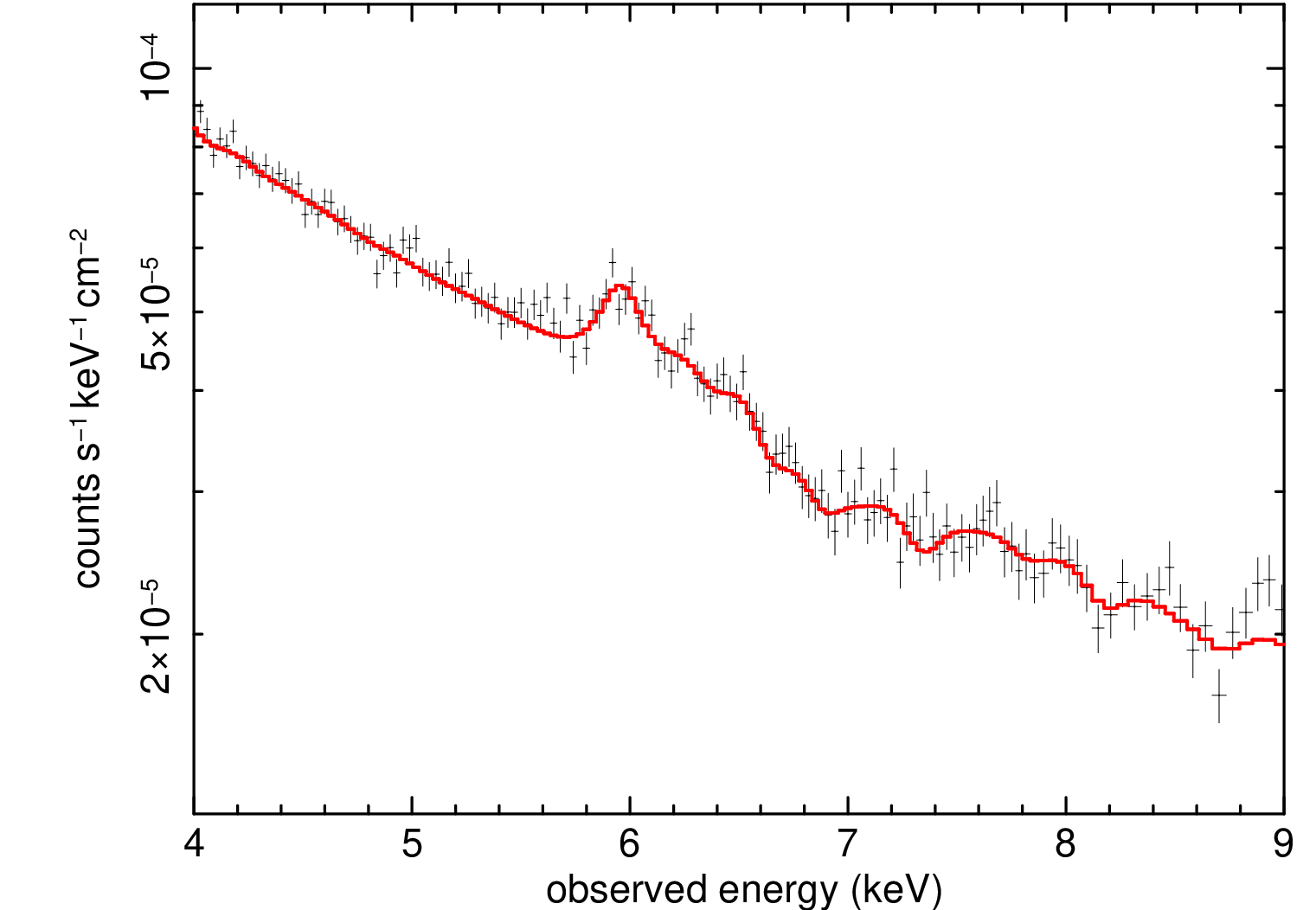}  
\caption{Comparison of stacked pn data and the final outflow model fit including a second highly ionized absorption component. (top) The principal emission 
and absorption lines in the two-component absorber model now provide a good match to spectral features in the data. Colour coding of the emission lines includes
black for the fluorescent Fe K line, and blue for the high-ionization emission. The hard power law and unabsorbed power
law components are shown in red and green respectively. Separate absorption line sequences of Lyman-$\alpha$, He-$\alpha$, Lyman-$\beta$ and He-$\beta$ for outflow velocities of $\sim$0.066c and
$\sim$0.129c are marked on the plot. (lower)  While lacking the visual clarity of a similar
plot obtained with higher resolution spectra data, individual emission and absorption features remain visible when folded through the CCD response function.}         
\end{figure}

\begin{table}
\centering
\caption{Final parameters of the highly ionized outflow obtained from a 2--10 keV spectral fit to the 2014 pn data, with two photoionized absorbers, defined by ionization parameter  $\xi$ (erg cm
s$^{-1}$), column density N$_{H}$ (cm$^{-2}$) and outflow velocity (v/c), together with a photoionized emission spectrum modelled by an ionization
parameter and outflow velocity. Extracted or added luminosities (erg s$^{-1}$) over the fitted spectral band 2--10 keV and the improvement in $\chi^{2}$ are also given for each photoionized model component}
\begin{tabular}{@{}lccccc@{}}
\hline
comp & log$\xi$ & N$_{H}$($10^{23}$)  & v/c & L$_{abs/em}$ & $\Delta \chi^{2}$ \\
\hline
abs & 4.0$\pm$0.2 & 3.7$\pm$2.9  & 0.129$\pm$0.002 & 5$\times10^{41}$ & 15/3 \\
abs & 3.4$\pm$0.1 & 2.0$\pm$1.0  & 0.066$\pm$0.003 & 1.8$\times10^{42}$ & 27/3 \\
emi & 3.5$\pm$0.1 & 1 (f) & 0.011$\pm$0.003 & 7$\times10^{41}$ & 34/3 \\
\hline
\end{tabular}
\end{table} 

From the dual velocity outflow spectral fit we find a mean 2-10 keV source luminosity of $\sim$6$\times 10^{43}$ erg s$^{-1}$.  The total extracted (absorbed) and added (emission) luminosities are 
$\sim$2.2$\times 10^{42}$ erg s$^{-1}$ and $\sim$7$\times 10^{41}$ erg s$^{-1}$, respectively, with the  lower velocity, lower ionization absorber contributing $\sim$80\% of the 2-10 keV opacity.  Table 1
summarises the main parameters of the photoionized absorption and emission spectra describing the highly ionized outflow in \pg\ in 2014.

To summarise, in XSPEC terms the final model is: TBabs(pl$_{soft}$ + (pl$_{hard})$(mtable)(mtable) + {\sc Xillver} + atable), where pl$_{hard}$ is the dominant continuum component ($\Gamma$$\sim$1.6), subject to
ionized absorption from the two outflow components. {\sc Xillver} represents the reflection continuum and associated Fe K fluorescence line, as described in the text.  pl$_{soft}$ is a soft continuum component
($\Gamma$$\sim$2.9) found from inter-orbit difference spectra where the subtraction of low flux spectra from high flux spectra show the residual spectrum is well described by a simple power law. While this variable, soft continuum component has little impact on the present analysis, it is included for consistency with the  soft x-ray analysis of \xmm\ grating spectra (Pounds \et\ 2015). 

Figure 4 (lower panel) compares the stacked 2014 pn data with the dual velocity outflow model (plotted immediately above), while Figure 5 shows the spectral model plot over the full 2--10 keV energy range to better 
illustrate  continuum and  line emission components. 

From the lower panel of Figure 5 we note the absence of a hard excess in the data-to-model ratio confirms that continuum reflection is adequately  modelled  by {\textsc Xillver}, at least up to 10 keV in our 
spectral fit, and does not indicate the strong reflection reported in Zoghbi \et (2015). Although discussed further at this point we note the strongest residual in the lower part of Figure 5 is excess emission
near 9 keV,  which - if confirmed - could represent blue-shifted FeXXVI RRC, perhaps indicative of a cooling flow.

\begin{figure}                                                          
\centering                                                              
\includegraphics[width=6.4cm, angle=270]{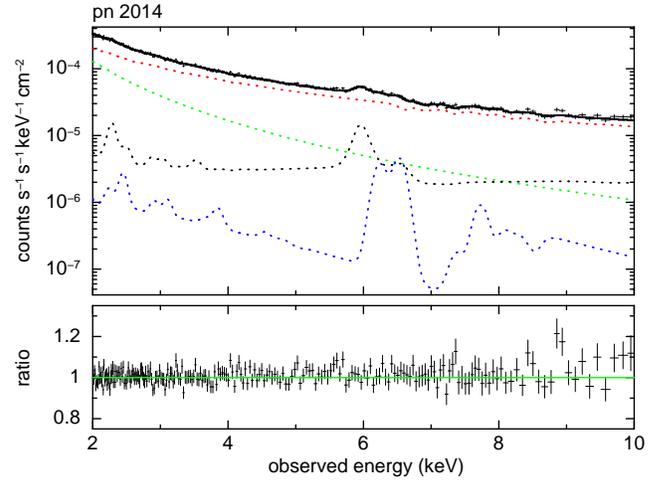}                                                 
\caption{(top) Comparison of the dual velocity outflow model and pn data over the full 2-10 keV band, showing the individual spectral components, with the power law and reflection continua (red, green and black), 
the fluorescent Fe K (also black) and ionized
emission lines (blue)}     
\end{figure}

\section{Supporting evidence for dual outflow velocities}

\subsection{Gaussian line fitting to the observed spectral structure}

Gaussian fitting to the emission and absorption line structure seen in the stacked pn data provides a model-independent check on the robustness of the above spectral analysis. The 4--10 keV ratio plot shown in 
Figure 6,  
obtained by removing the
photoionized absorption and emission spectra from the best fit model of Section 3, was scanned sequentially with
positive and negative Gaussians, having a minimum (1$\sigma$) line width of 50 eV comparable to the pn detector resolution. All features giving $\Delta \chi^{2}$
$\ga$6 were recorded and the sweep was then repeated with two broader absorption features at $\sim$7 keV and $\sim$8.2 keV re-fitted as narrow line pairs, making 9 absorption lines in all. The emission and absorption blend at $\sim$6.5--6.6 keV
was resolved by fixing the emission line width at 100 eV. The resulting fit is shown in Figure 6, with the measured line energies, equivalent width and proposed identifications listed in Table 2.  

Of the 9 narrow absorption lines in Table 2, those at $\sim$6.62, $\sim$6.87, $\sim$7.82 and $\sim$8.17 keV have observed line energies consistent with an outflow velocity in the range 0.065-0.069c when
identified with  the $\alpha$ and $\beta$ lines of FeXXV and FeXXVI. Taken as a line set, the weighted mean velocity is 0.067$\pm$0.001c and the combined significance is high ($\Delta \chi^{2}$ = 51/8),
providing strong model-independent support for the lower velocity absorber in Table 1. Four (of 5) remaining Gaussians, at $\sim$7.04, $\sim$7.31, $\sim$8.34 and $\sim$8.70 keV, are consistent with the
higher outflow velocity found in spectral modelling, with individual values - based on identification with the same set of  resonance absorption lines - ranging from 0.125-0.134c, and a mean value
v$\sim$0.128$\pm$0.002c. For this higher velocity line-set the statistical improvement of Gaussian line fitting is also highly significant  ($\Delta \chi^{2}$ =57/8).
Absorption line 5 does not fit the dual outflow velocity pattern, but is interesting in that the possible association with the FeXXV resonsnce line would match a third outflow velocity of v$\sim$0.19c
which gave a marginal improvement to the dual velocity spectral modelling described in Section 3.2.

\begin{figure}                                                          
\centering                                                              
\includegraphics[width=6.4cm, angle=270]{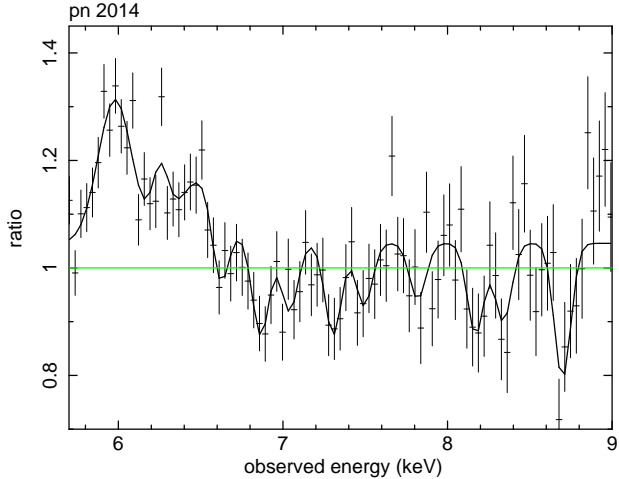}                                                 
\caption{Gaussian line fitting to spectral structure in the stacked pn data from the \xmm\ observation of \pg\ in 2014. In this plot data binning has been relaxed with the removal of the limit of 3 data points 
per resolution element to provide smoother Gaussian profiles. Nine possible absorption lines are detected, numbered from left to right as abs1 -- abs 9 in Table 2, where the measured energy, proposed identification and outflow velocity are
listed}     
\end{figure}

\begin{table*}
\centering
\caption{Sequential Gaussian fits to the positive and negative features in the pn 2014 data shown in Figure 6. Absorption lines (abs 1-9) have a fixed width comparable to the pn detector resolution, with the fitted line
energy, proposed identification and corresponding outflow velocity listed in each case.  Improvement in $\Delta \chi^{2}$ is after re-fitting the data at 4--10 keV following each added line.
All line identifications are with transitions in Fe XXV or Fe XXVI. The equivalent width of each line has been estimated separately in XSPEC}
\begin{tabular}{@{}lcccccccc@{}}
\hline
component & obs energy (keV) & EW (eV) &  rest energy (keV) & line id & energy (keV) & v/c & $\Delta \chi^{2}$ \\
\hline
Fe K-$\alpha$ & 5.98$\pm$0.01 & 80$\pm$7 & 6.46$\pm$0.01 & Fe K-$\alpha$ & 6.40 & 0.009$\pm$0.002 & 139/3\\ 
Fe 25  & 6.26$\pm$0.02 &  19$\pm$7 & 6.77$\pm$0.02 & He-$\alpha$ & 6.70 & 0.010$\pm$0.003 & 8/2\\
Fe 26 & 6.49$\pm$0.07 & 73$\pm$15 & 7.02$\pm$0.07 & Ly-$\alpha$ & 6.96 & 0.01$\pm$0.01 & 34/2\\
abs 1 & 6.62 $\pm$0.02 & -31$\pm$11  & 7.16$\pm$0.02 & He-$\alpha$ & 6.70 & 0.066$\pm$0.003 & 12/2 \\
abs 2 & 6.87$\pm$0.01 & -22$\pm$6  & 7.43$\pm$0.01 & Ly-$\alpha$ & 6.96 & 0.065$\pm$0.002 & 16/2\\
abs 3 & 7.04 $\pm$0.02 & -11$\pm$7 &  7.61$\pm$0.02 & He-$\alpha$ & 6.70 & 0.127$\pm$0.003 & 7/2\\
abs 4 & 7.31$\pm$0.02 & -22$\pm$8 & 7.90$\pm$0.02 & Ly-$\alpha$ & 6.96 & 0.126$\pm$0.003 & 25/2 \\
abs 5 & 7.49$\pm$0.03 & -9$\pm$7 & 8.10$\pm$0.03 & He-$\alpha$ & 6.70 & 0.188$\pm$0.004 & 7/2 \\ 
abs 5 & 7.49$\pm$0.03 & -9$\pm$7  & 8.10$\pm$0.03 & Ly-$\alpha$ & 6.96 & 0.151$\pm$0.004 & 7/2 \\ 
abs 6 & 7.82$\pm$0.03 & -16$\pm$9 & 8.45$\pm$0.03 & He-$\beta$ & 7.88 & 0.069$\pm$0.003  & 10/2 \\ 
abs 7 & 8.17$\pm$0.02 & -20$\pm$8   & 8.83$\pm$0.02 & Ly-$\beta$ & 8.25 & 0.068$\pm$0.002 & 13/2\\ 
abs 8 & 8.34$\pm$0.04 & -16$\pm$9  & 9.02$\pm$0.04 & He-$\beta$ & 7.88 & 0.134$\pm$0.005 & 6/2 \\ 
abs 9 & 8.70$\pm$0.02 & -41$\pm$11   & 9.40$\pm$0.02 & Ly-$\beta$ & 8.25 & 0.129$\pm$0.003 & 19/2\\ 
abs 9 & 8.70$\pm$0.02 & -41$\pm$11  & 9.40$\pm$0.02 & He-$\gamma$ & 8.29 & 0.125$\pm$0.003  & 19/2 \\ 
\hline
\end{tabular}
\end{table*}

\section{Discussion}

The extended observation of \pg\ in 2014 has provided high quality hard x-ray spectra revealing previously unseen spectral structure of a UFO in the $\sim$6-10 keV energy band. Spectral modelling
has identified the observed absorption structure with resonance and higher order lines  of highly ionized Fe, consistent with two distinct outflow velocities. The faster wind component has a velocity of
v$\sim$0.13c, similar to that first seen in 2001, but with a higher column density and ionization parameter.  The second  outflow component, not resolved in the EPIC data in 2001, has a similar column
density but a lower velocity v$\sim$0.066c.

Key factors in identifying the lower velocity flow in 2014 were the high quality of the stacked 2014 spectral data, allowing spectral features to be identified near the limit of CCD energy resolution, and
the simultaneous modelling of highly ionized emission, which allowed - in particular - the lower velocity Fe XXV He-$\alpha$ absorption line to be detected.   The photoionized  emission spectrum also
successfully reproduces the higher energy features observed in the stacked 2014 pn data, identified with resonance emission from He- and H-like Fe ions in a similar ratio to that seen in absorption.
Comparison of absorbed and (re-) emission luminosities indicates a substantial covering factor of the highly ionized wind.

An important outcome of the present analysis is in finding that the ultra-fast  outflow discovered in an \xmm\ observation in 2001 is again detected in 2014. Although not resolved in the Fe K spectrum in
2001, there is independent evidence for the co-existence of the lower velocity outflow in that earlier observation, from a re-analysis of the soft x-ray spectrum (Pounds 2014b) and in an earlier
partial-covering spectral fit which required an absorber moving at $\sim$0.07c to explain continuum curvature (Pounds and Reeves 2009).

The properties of powerful AGN winds are reviewed in King and Pounds (2015), where a highly ionized wind is envisaged being launched at the local escape velocity by continuum photons from a SMBH accreting
at modest Eddington ratios  $\dot m = \dot M/\me \sim 1$, finding  - for  accretion from a disc - the excess accreting matter is expelled in a quasi-spherical wind, with a launch velocity $v \simeq
{\eta\over \dot m}c \sim 0.1c $. Observational support for that picture is provided by recent archival searches (Tombesi \et\ 2010, 2011; Gofford \et\ 2013) which find a substantial fraction of luminous
AGN having a highly ionized wind with a velocity in the range $\sim$0.03--0.3c 

The new observations indicate a more complex picture, with two distinct outflow velocities, co-existing in observations 13 years apart. Chaotic accretion (King and Pringle 2006), consisting of many prograde
and retrograde events, offers an intriguing explanation of such a dual velocity wind, the persistence of two distinct outflow velocities perhaps relating to physically distinct orientations of the inner
accretion flow, both close to Eddington, and with differing values of the accretion efficiency $\eta$ and hence of velocity. That possibility is discussed in more detail by King and Nixon (2016)

Higher resolution hard x-ray spectra from the forthcoming \astroh\ observatory should show how common are such complex highly ionized AGN winds as reported here for \pg.

\section{Conclusion}

An extended \xmm\ observation of the luminous narrow line Seyfert galaxy \pg\ in 2014 has revealed previously unseen spectral structure in Fe K absorption, finding a second high velocity
component of the highly ionized wind. In identifying that additional complexity within the limits of CCD energy resolution, the 2014 observation benefited critically from the high statistical significance of the EPIC data
resulting from the unusually long observation. The outcome promises further revelations of the dynamical structure of AGN winds in a new era of high resolution hard x-ray spectroscopy, heralded with the
near-future launch of \astroh. 

\section*{Acknowledgements}

\xmm\ is a space science mission developed and operated by the European Space Agency. We acknowledge the excellent work of ESA staff in Madrid in 
planning and conducting the \xmm\ observations. The UK Science and Technology Facilities Council 
funded the postdoctoral research assistantship of AL. We acknowledge the continued cooperation with the Leicester theory group led by Andrew King. The present text has benefited significantly in clarity from
several constructive suggestions from the anonymous referee.

\end{document}